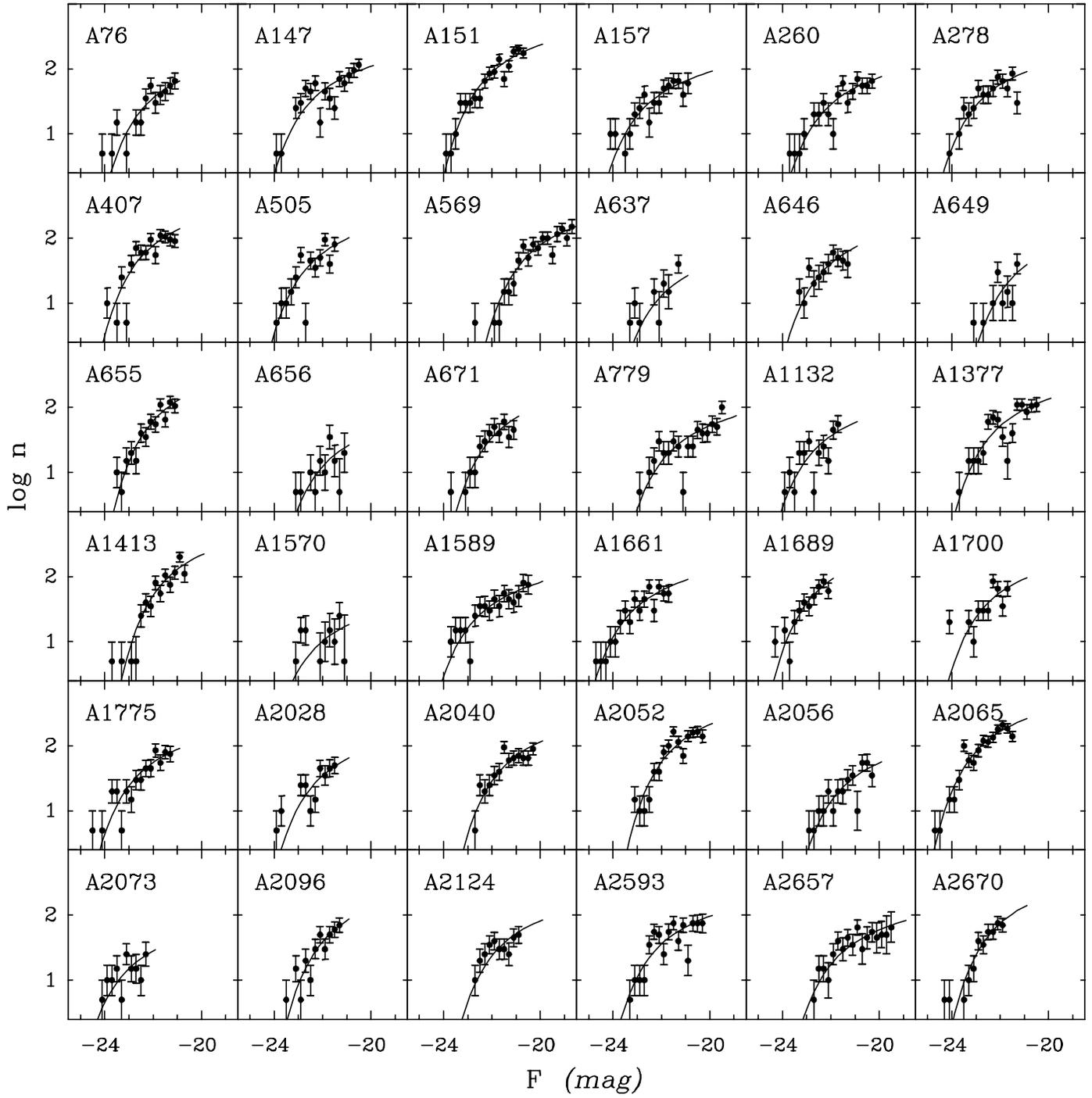

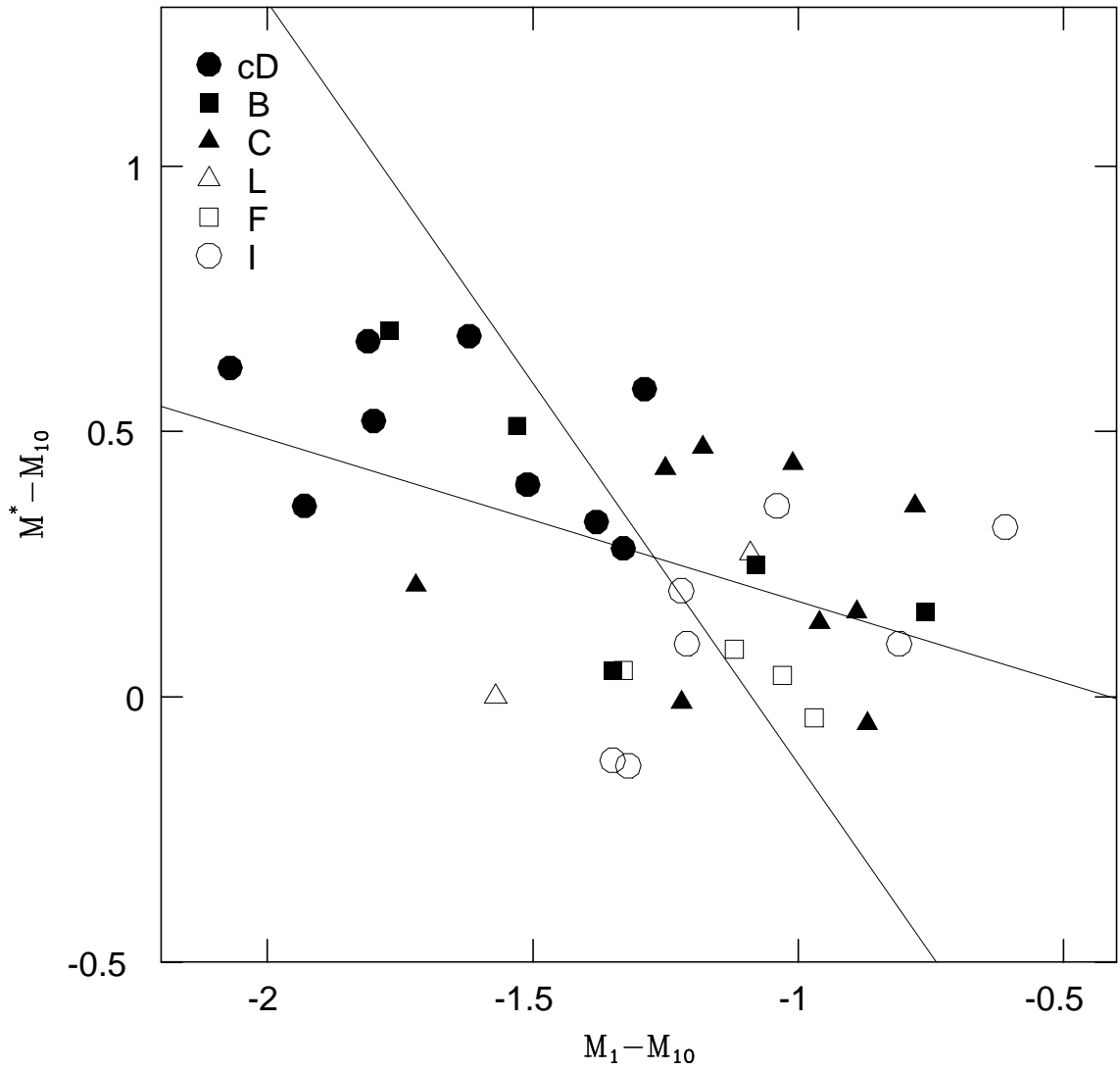

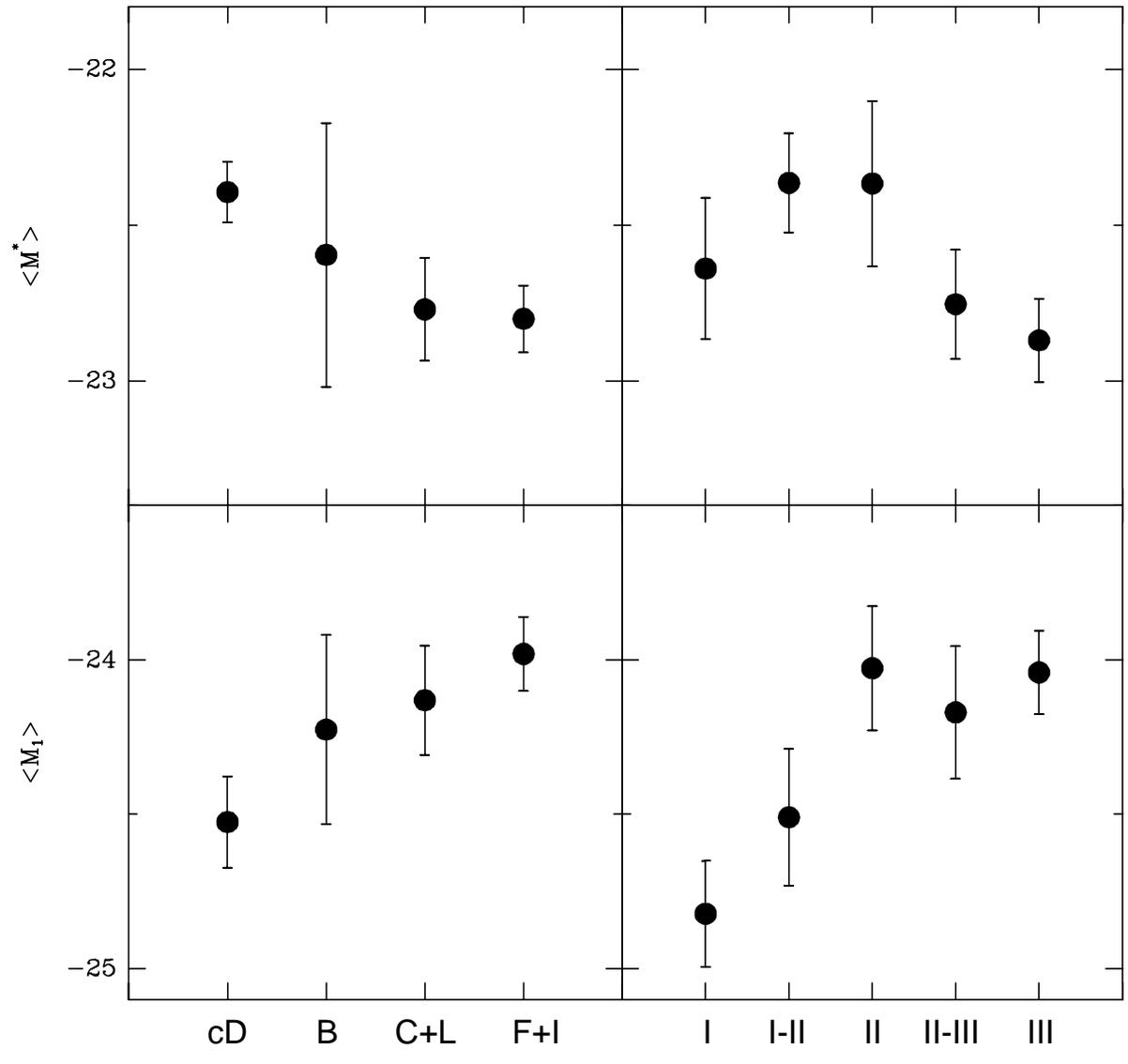

# The luminosity function of cluster galaxies: relations among $M_1$, $M^*$ and the morphological type

D. Trèvese, G. Cirimele, and B. Appodia

Istituto Astronomico, Università di Roma "La Sapienza", via G. M. Lancisi 29, I-00161 Roma, Italy

**Abstract.**
A study of the luminosity function of 36 Abell clusters of galaxies has been carried out using photographic plates obtained with the Palomar 1.2 m Schmidt telescope. The relation between the magnitude $M_1$ of the brightest cluster member and the Schechter function parameter $M^*$ has been analyzed. A positive correlation between $M^*$ and $M_1$ is found. However clusters appear segregated in the $M_1$-$M^*$ plane according to their Rood & Sastry class in such a way that on average $M_1$ becomes brighter while $M^*$ becomes fainter going from late to early Rood & Sastry and also Bautz & Morgan classes. Also a partial correlation analysis involving the magnitude $M_{10}$ of the 10th brightest galaxy, shows a negative intrinsic correlation between $M_1$ and $M^*$. These results agree with the cannibalism model for the formation of brightest cluster members, and provide new constraints for theories of cluster formation and evolution.

## 1. Introduction

The observed luminosity function (LF) of galaxies provides the most important means to test theories of galaxy formation and evolution. According to a reformulation and extension of the Press & Schechter (1974) theory presented by Bond et al. (1991), the direct hierarchical clustering of primordial overdensities is responsible for the onset and evolution of a self-similar mass distribution leading to the observed Schechter-like galaxy luminosity function.

Moulding of the luminosity function by galaxy merging has recently been exploited (Cavaliere & Menci 1993 and refs. therein) to reconcile the local LF (Efstathiou, Ellis & Peterson 1988; Binggeli, Sandage & Tamman 1988) with faint galaxy counts (Tyson & Seitzer 1988) and redshift distribution (Broadhurst, Ellis & Shanks 1988).

Galaxy merging, or galactic cannibalism, was originally introduced by Ostriker and Tremaine (1975) as a mechanism of growth of brightest cluster members at the expense of the other massive galaxies, which are most affected by dynamical friction. According to the model of Hausman and Ostriker (1978), as the evolution of the cluster proceeds this selective depletion should push to lower luminosities the turnover point between the steep high-luminosity falloff and the flatter faint end of the LF. This should cause a negative correlation between the magnitude $M_1$ of the brightest cluster member and the characteristic magnitude $M^*$ of a fitting Schechter (1976) LF.

From a study of 12 rich clusters, Dressler (1978) derived an indication that $M_1$ and $M^*$ are anticorrelated. However, in a subsequent study of 9 Abell clusters, Lugger (1986) (L86) concluded that the correlation is not statistically significant and, in a more recent study of 12 Abell clusters, Oegerle and Hoessel (1989) (OH89) find no evidence for any relation between $M_1$ and $M^*$. These results are taken, by the respective authors, as indications against the effect found by Dressler (1978). It should be noted that in both these studies a positive $M_1$-$M^*$ correlation is found, though it is not statistically significant due to the small number of clusters analyzed.

Since the properties of individual clusters are strongly affected by random fluctuations, they become meaningful only when defined statistically using a large number of clusters. Thus a project has been undertaken for studying in a uniform manner a large sample of nearby galaxy clusters (Flin et al. 1988, Trèvese et al. 1992 (T92), Flin et al. 1995), and deriving properties such as number density profiles, morphology, galaxy orientations, luminosity functions (LF) and their possible statistical relations.

The results concerning galaxy orientations in a sample of 55 clusters (Trèvese, Cirimele and Flin 1992) indicate the special role played by first ranked galaxies during the evolution of clusters.

In the present paper we restrict our attention to the specific problem of the $M_1$-$M^*$ relation and present the results obtained from a sample of 36 Abell clusters, more than three times larger than each of the previous samples. From this we obtain a statistically significant evidence of a new type of negative $M_1$-$M^*$ partial correlation, related to the fact that $M_1$ becomes brighter and $M^*$ fainter in going from irregular to cD Rood & Sastry (1971) (RS) cluster types or from type III to type I in the Bautz & Morgan (1970) (BM) classification.

## 2. Data and Reductions

**Table 1.** Cluster Data

| Abell # | z | $M_1$ | $M_{10}$ | $M^*$ | RS | BM | N |
|---|---|---|---|---|---|---|---|
| A76[a] | 0.0416 | -24.30 | -22.73 | -22.73 | L | II-III | 77 |
| A147[a] | 0.0438 | -24.12 | -22.91 | -22.81 | I | III | 152 |
| A151[a] | 0.0536 | -24.88 | -23.26 | -22.58 | cD | II | 260 |
| A157[a,1] | 0.103[†] | -24.54 | -23.19 | -23.14 | B | II | 106 |
| A260[b] | 0.0348 | -23.96 | -22.63 | -22.58 | F | II | 109 |
| A278[b,1] | 0.0896 | -24.09 | -23.28 | -23.18 | I | III | 114 |
| A407[c] | 0.0463 | -24.11 | -23.07 | -22.71 | I | II | 168 |
| A505[c] | 0.0543 | -24.57 | -23.19 | -22.86 | cD | I | 94 |
| A569[c] | 0.0196 | -23.39 | -21.62 | -20.93 | B | II | 229 |
| A637[d,1] | 0.136[†] | -23.85 | -22.63 | -22.62 | | | 23[#] |
| A646[d,1] | 0.1303 | -23.85 | -23.24 | -22.92 | I | III | 74[#] |
| A649[d,1] | 0.124[†] | -23.73 | -22.40 | -22.12 | cD | II | 25[#] |
| A655[d] | 0.1245 | -25.01 | -23.20 | -22.53 | cD | I-II | 131 |
| A656[d,1] | 0.136[†] | -23.77 | -22.42 | -22.54 | I | III | 33[#] |
| A671[d] | 0.0502 | -24.29 | -22.57 | -22.36 | C | II-III | 71 |
| A779[d] | 0.0230 | -24.44 | -22.64 | -22.12 | cD | I-II | 108 |
| A1132[e] | 0.1363 | -23.93 | -23.17 | -23.01 | B | III | 51[#] |
| A1377[d] | 0.0514 | -24.06 | -22.98 | -22.73 | B | III | 175 |
| A1413[d] | 0.1427 | -25.01 | -22.94 | -22.32 | cD | I | 172 |
| A1570[d,2] | 0.156[†] | -24.03 | -22.71 | -22.84 | I | II-III | 20[#] |
| A1589[d] | 0.0718 | -24.22 | -23.33 | -23.17 | C | II-III | 122 |
| A1661[f] | 0.1671 | -24.83 | -23.71 | -23.62 | F | III | 101 |
| A1689[d] | 0.1832 | -25.04 | -23.86 | -23.39 | C | II-III | 84[#] |
| A1700[f,1] | 0.119[†] | -24.28 | -23.19 | -22.92 | L | III | 78[#] |
| A1775[d] | 0.0717 | -25.21 | -23.68 | -23.17 | B | I | 107 |
| A2028[d] | 0.0776 | -24.20 | -22.98 | -22.78 | I | II-III | 53 |
| A2040[d] | 0.0456 | -23.20 | -22.42 | -22.06 | C | III | 132 |
| A2052[d] | 0.0348 | -23.95 | -22.66 | -22.08 | cD | I-II | 253 |
| A2056[d,3] | 0.0804 | -23.11 | -22.15 | -22.01 | C | II-III | 64[#] |
| A2065[d] | 0.0722 | -24.80 | -23.79 | -23.35 | C | III | 291 |
| A2073[d,3] | 0.1717 | -24.37 | -23.50 | -23.55 | C | III | 27[#] |
| A2096[d,4] | 0.108[†] | -23.98 | -22.73 | -22.30 | C | III | 69[#] |
| A2124[d] | 0.0654 | -24.50 | -22.57 | -22.21 | cD | I | 62[#] |
| A2593[g] | 0.0421 | -23.58 | -22.55 | -22.51 | F | II | 137 |
| A2657[a] | 0.0414 | -23.25 | -22.28 | -22.32 | F | III | 125 |
| A2670[d] | 0.0761 | -24.64 | -23.13 | -22.73 | cD | I-II | 74 |

[†] the redshift has been estimated from $z$-$m_{10}$ relation.
Zero of photometric scale from: [a] Hoessel, Gunn & Thuan 1980; [b] Sandage & Perelmuter 1991; [c] Peterson 1970; [d] Hoessel & Schneider 1985; [e] Gunn & Oke 1975; [f] Bothun et al., 1985; [g] Murphy, Schild & Weekes 1983; [a1] [b1] [f1] same plate as A151, A260, A1661 respectively. [d1] [d2] [d3] [d4] same plate as A655, A1589, A2065 A2124 respectively. [#] magnitude limits brighter than $m_3 + 3$.

Our data are derived from 10-inch photographic plates taken by P. Hickson with the 48-inch Palomar Schmidt Telescope to analyze a sample of 64 Abell clusters. The fields were selected according to the criteria specified in Hickson (1977) together with details about the emulsions and filters employed. The resulting photometry corresponds to the red F-band of Oemler (1974). The sample was not statistically complete but it was designed to cover all cluster morphological types, to perform statistical studies of each cluster type. Other clusters appearing in the same plates, some of which are not Abell clusters but belong to the Zwicky catalog, were also added to the sample, thereby reaching a total of more than 100 clusters in all. Plates were scanned with a PDS 1010G in Rome, with pixel sizes ranging from 10 to 25 $\mu m$ depending on the cluster distance. Automatic identification of objects and their classification as point-like or diffuse are described in T92. Total magnitudes are computed from the flux integrated in a circular aperture whose radius is $R_1 = 1.5 r_1$, where $r_1$ is the first moment of the intensity distribution (see T92). The magnitude defined in this way corresponds on average to an isophotal magnitude at 24 $mag \cdot arcsec^{-2}$, with the advantage that $r_1$ is less noisy than the corresponding isophotal radius. The signal to noise ratio is $S/N \gtrsim 100$ for a few objects brighter than $F = 12$ $mag$, about 25 for $F \approx 14$ $mag$ and falls to about 5 for $F \gtrsim 18$ $mag$.

Relative photometry has been obtained for 55 of the above clusters (Trèvese, Cirimele and Flin 1992), while the zero of the magnitude scale has been established for 36 clusters, using published photometric data. For 27 out of 36 clusters we used $r$ band data from Hoessel, Gunn & Thuan (1980) and Hoessel & Schneider (1985); for 8 of the remaining clusters we used $V$ data while for A2593 we used $R$ data, as specified in the table.

In the case of Johnson R-band data we assume F=R as discussed in Lugger (1989). For data in the Thuan and Gunn (1976) $r$ and Johnson $V$ we assume the average color of bright cluster galaxies implying $F = r - 0.58$ and $F = V - 0.76$, as given by Schneider et. al. (1983) from which we also take the relevant k-corrections. Magnitudes have been corrected for the interstellar extinction $A_F$ by adopting the relation $A_F = 0.07(\csc b - 1)$ (Oemler 1974). Overall, the estimated magnitude error due to the uncertainty of the zero point of each plate plus the internal error is less than 0.2 $mag$.

## 3. Luminosity function determination

To compare the luminosity functions of different clusters it is important to define the galaxy samples of individual clusters according to uniform criteria, as discussed in L86. In particular, the distribution of galactic types and luminosities varies from the cluster core to the field (Oemler 1974, Lugger 1986). Thus the luminosity functions were determined inside circular regions with a fixed radius of $R_A = 1.7/z$ $arcmin$ corresponding to 3 Mpc for $H_o$=50 $Km$ $s^{-1}$ $Mpc^{-1}$, $q_o = 1$. Outside these regions a local field density was computed and compared with the field galaxy counts of Butcher & Oemler (1985). The agreement was within 10 percent in 23 cases. Though the local determination is probably more appropriate, for the statistical comparison of different clusters we preferred to adopt the average background to treat all the clusters in a uniform way and to obtain a closer comparison with the work of previous authors. In particular the background counts as a function of the red apparent magnitude $m_R$ were deduced from Butcher & Oemler (1985) assuming R=F and an average color index J-F=1.0 as in Lugger (1989). A a straight line fit in the range $14 \leq m_R \leq 18$ gives $\log N_b = 0.503 \cdot m_R - 7.49$, where $N_b$ is the number of background galaxies per square degree and per 0.25 magnitudes interval.

The galaxy samples of each cluster were corrected statistically by eliminating, in each magnitude bin of 0.2 $mag$, a number of galaxies estimated from the field density in the same magnitude bin. Finally a magnitude limit as close as possible to $m_3 + 3$ was adopted in most (24) cases while for the remaining 12 clusters, whose fainter magnitudes bins have an anomalously low population, a brighter limit was adopted (these objects are marked with # in column 8 of Table 1).

The LFs were then fitted with a Schechter (1976) function:

$$\Phi(L)dL = \Phi^* \left(\frac{L}{L^*}\right)^\alpha exp\left(-\frac{L}{L^*}\right) d\left(\frac{L}{L^*}\right) \qquad (1)$$

maximizing the likelihood, as in Schechter & Press (1976), by means of the MINUIT package of the CERN library. Preliminary results concerning a non parametric comparison of the luminosity functions have been presented elsewhere (Trèvese et al. 1996).

Since our aim is to analyze the relation between $M^*$ and the magnitude $M_1$ of the brighter cluster member, each LF has been fitted with $M^*$ as free parameter and $\alpha$ fixed to the universal value −1.25 (Schechter 1976) and the first ranked galaxy has been excluded from the fit (see OH89).

...

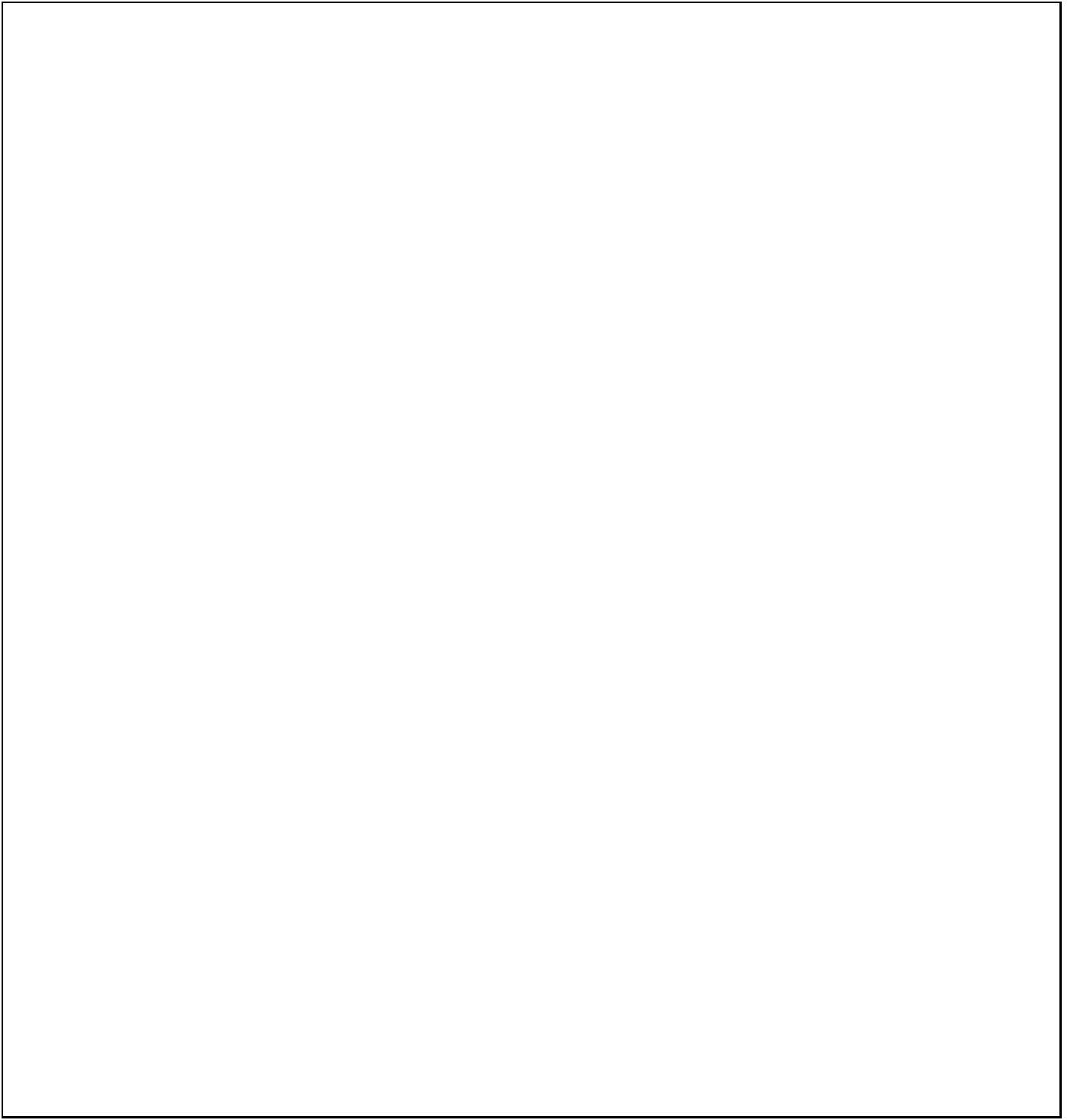

**Fig. 1.** The luminosity functions in the $F$ band of the 36 clusters of the sample. Each point represents the logarithm of the number $n$ of galaxies in bins of 0.2 *mag*. The continuous curves represent the maximum likelihood fit with a Schechter function of the unbinned data. The brightest galaxy is not included in the fit and is not shown.

The resulting $M^*$ values for the entire sample of 36 clusters are reported in column 5 of Table 1 together with the Abell catalogue number, redshift (Struble & Rood 1991), $M_1$, $M_{10}$, Rood & Sastry and Bautz & Morgan classes, and the number of galaxies of each cluster used in the fitting. The source of morphological classification are Struble & Rood (1987) for the RS type and Abell, Corwin & Olowin (1989) for BM types. The differential LFs are shown in Figure 1 with the fitting function. It is to be noted however that the curve shown is not a fit to the points in the figure, which derive from an arbitrary binning of the data, but represents the maximum likelihood fit to the unbinned data.

## 4. The statistics of $M_1$ and $M^*$

The values of $M^*$ show a nearly gaussian distribution with $M^* = -22.66 \pm 0.52$ in agreement with the values $-22.52 \pm 0.45$, $-22.64 \pm 0.50$, $-22.85 \pm 0.23$, found respectively by Dressler (1978), L86 and OH89.

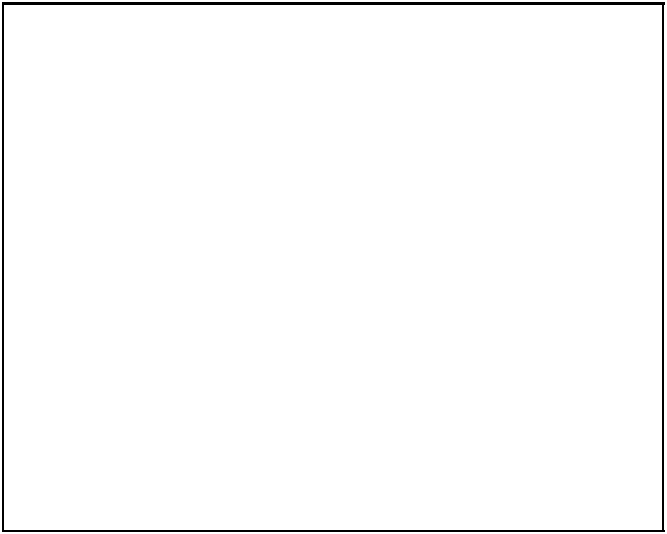

**Fig. 2.** $M^*$ versus $M_1$ for the 36 clusters of the sample. The different symbols refer to the RS classes, as indicated. A positive global correlation is seen.

Figure 2 shows the distribution of the 36 clusters of our sample in the $M_1$-$M^*$ plane. A fit of a straight line $M^* = a_o + a_1 M_1$ to the data, assuming equal errors on both axes, gives $a_1 = 1.12$, or $a_1 = 0.58$ considering only the errors on $M^*$. The correlation is positive with a coefficient $r = 0.54$ and an associated probability $P(>r) = 7 \cdot 10^{-4}$ of the null hypothesis. The clusters appear segregated according to their RS class, with early types towards the top left and late types towards the bottom right.

Notice that L86, from a sample of 9 clusters, obtains $a_1 = 0.67 \pm 0.37$, $r = 0.56$ corresponding to $P(>r) = 12\%$, while from the data of OH89, relative to 12 clusters, it is possible to derive $a_1 = 0.22 \pm 0.14$, $r = 0.45$ corresponding to $P(>r) = 14\%$. Thus our result gives a statistically significant proof of the trend suggested by the previous findings of L86 and OH89. These results contrast with the findings of Dressler (1978) who obtains, for 12 clusters, $a_1 \approx -0.5$, $r \approx -0.5$, but a much higher probability of the null hypothesis due to the smaller size of the sample (see the discussion of L86 about the role played by A665 which has a somewhat uncertain value of $M^*$).

A straightforward interpretation would be that the selective depletion of the bright end of the luminosity function, predicted by the galactic cannibalism model of Hausman and Ostriker (1978), does not agree with the observations.

However, before deriving any physical conclusion from the observed correlation, it must be considered that the error on the zero point of the magnitude scale affects by the same amount both $M_1$ and $M^*$ of the same cluster, causing a positive $M_1$ - $M^*$ correlation, even in the case the true values of $M_1$ and $M^*$ are intrinsically uncorrelated. It is easy to show that, in the absence of any intrinsic $M_1$-$M^*$ correlation, the observed correlation coefficient would be $r_o = \sigma_c^2/(\sigma_1 \cdot \sigma^*)$, where $\sigma_1$ and $\sigma^*$ are the observed standard deviations of $M_1$ and $M^*$ respectively, and $\sigma_c$ is the standard deviation of the calibration error (see Massaro and Trèvese (1996)). Thus, in our case a r.m.s. uncertainty e.g. as large as 0.3 $mag$ on the calibration could account for the observed positive correlation. However, in a few cases it has been possible to find in the literature other photometric data on the galaxies we used to establish the zero point, obtained by different authors, sometimes in different bands. A comparison of these data shows that the uncertainty is less than 0.2 $mag$. Thus it is possible that the calibration uncertainty accounts for the positive $M_1$-$M_*$ correlation only in part. In any case a reliable correction of this statistical bias would require an accurate estimate of both the random photometric noise and the calibration errors.

A possible approach consists in studying magnitude differences like $(M^* - M_{10})$ and $(M_1 - M_{10})$, which are independent of calibration errors, and of any global shift of the luminosity function. The choice of the 10-th ranked galaxy is motivated by the fact that the r.m.s. deviation of $M_k$ has a minimum for $k = 10$ in our sample of 36 clusters.

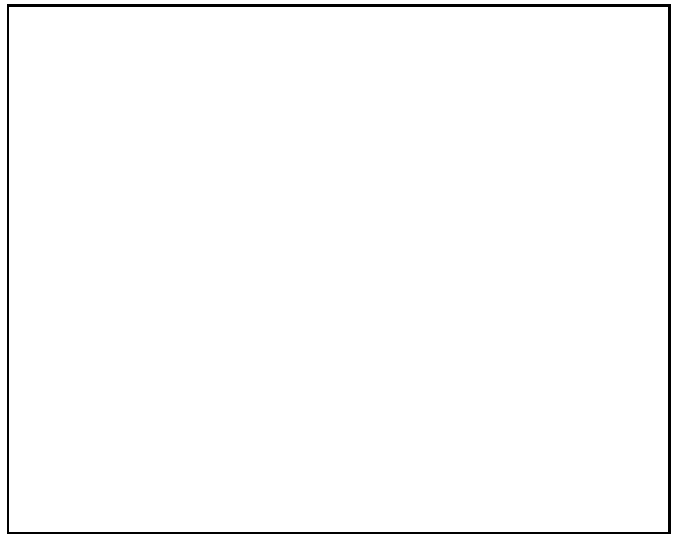

**Fig. 3.** $(M^* - M_{10})$ versus $(M_1 - M_{10})$ for the 36 clusters of the sample. The different symbols refer to the RS classes, as indicated. A negative correlation is seen.

$(M^* - M_{10})$ is plotted versus $(M_1 - M_{10})$ in Figure 3 and appears negatively correlated: $r = -0.46$, $P(>|r|) = 4.6 \cdot 10^{-3}$. The clusters are still segregated according to their RS classes.

Alternatively it is possible to perform a partial correlation analysis in order to single out the intrinsic correlation between $M_1$ and $M^*$, removing the correlation produced by a global shift of the luminosity scale.

In the case of three (or more) stochastic variables $x_i$, $i = 1,2,3$, from the ordinary (zero order) correlation coefficients $r_{i,j}$, it is possible to compute the partial correlation coefficients $r_{i,j;k}$, $(i,j,k = 1,2,3; i \neq j \neq k)$ defined by:

$$r_{i,j;k} = \frac{r_{i,j} - r_{j,k} \cdot r_{i,k}}{\sqrt{(1 - r_{j,k}^2) \cdot (1 - r_{i,k}^2)}}, \qquad (2)$$

which represents the *intrinsic* correlation between $x_i$ and $x_j$ corrected for the effect induced by the correlation of both $x_i$ and $x_j$ with $x_k$ (see e.g. Anderson 1984).

The ordinary correlation coefficients between $M_1$, $M^*$ and $M_{10}$ are all positive: $r_{1,*}=0.54$, $r_{1,10}=0.76$ and $r_{*,10}=0.89$. The resulting partial correlation coefficient is $r_{1,*;10} = -0.50$ with an associated probability $P(>|r|) = 1.9 \cdot 10^{-3}$.

The effect is statistically significant, thus providing a new constraint for any model of cluster formation and evolution.

As already pointed out, in Figures 2 and 3 the clusters are segregated according to their RS class. The same effect appears using the BM classes. Thus we have divided the clusters into groups, corresponding to the RS classes F+I, C+L, B and cD respectively, to collect enough objects in each group. Then, for comparison, we have also grouped the clusters according to their BM class.

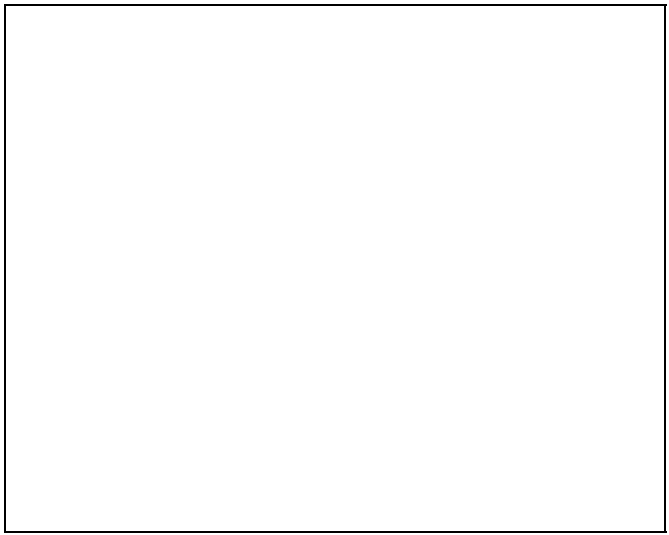

**Fig. 4.** The average values $<M_1>$ (lower panels) and $<M^*>$ (upper panels), computed for the subsamples corresponding to the Rood & Sastry classes cD, B, C+L and F+I (left) and to the Bautz & Morgan types (right). From late to early types $<M_1>$ becomes brighter, while $<M^*>$ becomes cluster fainter.

The average values $<M_1>$ and $<M^*>$, taken over each group, result negatively correlated. The effect is better seen in Figure 4, where $<M_1>$ and $<M^*>$ of the different groups are shown. It appears that $<M_1>$ becomes brighter, while $<M^*>$ becomes fainter, in going from irregular to compact and regular clusters.

A more precise characterization of this trend is obtained by computing, for the 36 clusters of our sample, the Kendall rank correlation coefficients $\tau$ of $M_1$, $M^*$ and $M_{10}$ with the RS and BM type, together with the relevant probabilities $P(\tau)$ of the null hypothesis. The results are summarized in Table 2.

A negative correlation appears of $M_{10}$ with RS and BM types, but it is very weak and not statistically significant. It is possible to check from data in Table 1, that an error in the color transformation of 0.2 *mag*, e.g. causing a systematic shift of 1/4 of the sample, does not appreciably change the correlation coefficients and the relevant probabilities.

We also repeated the statistical calculations excluding from the sample the 12 clusters whose LFs have been determined in smaller range of magnitude. The results are essentially unchanged (in some cases the significance of correlation in even slightly improved).

As a result we can conclude that we obtained for the first time significant evidence of a negative correlation of $M^*$ with both the RS and BM classes. We find also that the correlations of $M^*$-$M_{10}$ and $M_1$-$M_{10}$ with the morphological classes are the most significant. Finally we find a trend of $M_1$ with the RS cluster type and we confirm the similar trend of $M_1$ with the BM cluster type which was already known (see e.g. Sandage & Hardy 1973).

## 5. Conclusions

We can summarize our results as follows:

- The mean of the characteristic magnitudes $M^*$, determined by maximum likelihood fits, is in good agreement with the values found in the literature.
- $M^*$ is positively correlated with $M_1$ with high statistical significance.
- Clusters appear segregated in the $M_1$-$M^*$ plane according to both their Rood & Sastry and Bautz & Morgan type.
- Including in the study also the magnitude $M_{10}$ of the 10th brightest member, a partial correlation analysis shows a negative intrinsic correlation between $M_1$ and $M^*$.
- We find statistically significant evidence that on average the magnitude $M_1$ of the brightest cluster member is brighter in clusters of the earlier Rood & Sastry and Bautz & Morgan types.
- The characteristic magnitude $M^*$ is on average fainter in clusters of the earlier Rood & Sastry and Bautz & Morgan types. The effect is statistically significant, providing a new constraint for theories of cluster formation and evolution.

Once the RS types are interpreted as an evolutionary sequence going from late type, irregular clusters to the more concentrated, dynamically evolved cD clusters, the above results may support the cannibalism model of Hausman and Ostriker (1978). In this scheme, during the cluster evolution the first ranked galaxies grow by merging, becoming brighter.

We stress that merging itself is not sufficient to explain the increase of $M^*$. Rather one must assume that the merging affects preferentially the most massive and bright cluster members. This also causes the increase of $M_{10}$, which however is found to be smaller. It is also important to remember that $M_1$ is excluded from the maximum likelihood fit, as done by previous investigators, so that the determination of $M^*$ is not affected by the brightening of $M_1$ along the RS sequence, from I to cD types, but simply measures the depletion of the bright end of the luminosity function.

The emerging scenario could be the following. The positive global $M_1$-$M^*$ correlation which appears in the data is mostly intrinsic. Possibly because clusters are born with different luminosity functions which, to a first approximation, are globally shifted towards brighter or fainter luminosities, according to some statistical distribution. During the subsequent evolution the brightest member grows at the expense of other bright galaxies, which are most affected by dynamical friction, causing the depletion of the bright end of the LF. This happens for both intrinsically brighter and fainter clusters. Once the spread in luminosity is reduced by computing the average over individual classes, the negative $M_1$-$M^*$ correlation appears.

**Table 2.** Rank correlation statistics

|         | RS     |                | BM     |                |
|---------|--------|----------------|--------|----------------|
|         | $\tau$ | $P(\tau)$      | $\tau$ | $P(\tau)$      |
| $M_1$       | 0.26   | $2.3 \cdot 10^{-2}$ | 0.30   | $9.8 \cdot 10^{-3}$ |
| $M^*$       | $-0.24$ | $4.0 \cdot 10^{-2}$ | $-0.25$ | $3.0 \cdot 10^{-2}$ |
| $M_{10}$    | $-0.03$ | $9.7 \cdot 10^{-1}$ | $-0.05$ | $6.2 \cdot 10^{-1}$ |
| $M_1$-$M_{10}$ | 0.37 | $1.2 \cdot 10^{-3}$ | 0.58 | $5.4 \cdot 10^{-7}$ |
| $M^*$-$M_{10}$ | $-0.47$ | $4.4 \cdot 10^{-5}$ | $-0.34$ | $3.7 \cdot 10^{-3}$ |

We stress that, beyond any evolutionary interpretation, our results show that the cluster LFs are not universal but depend on the cluster type in a systematic way.

Intrinsic differences and evolutionary changes of the shape of the LF will be better understood through a non parametric analysis.